\documentclass[twocolumn,pre,floatfix,superscriptaddress]{revtex4}
\usepackage{graphicx}
\usepackage{amsmath}
\usepackage{latexsym}
\usepackage{epstopdf}
\usepackage{amsmath}
\usepackage{amssymb,amsmath}

\newcommand{\beq}{\begin{equation}}
\newcommand{\eeq}{\end{equation}}

\begin{document}

\title{Contour lines of the discrete scale invariant rough surfaces}
\author{M. Ghasemi Nezhadhaghighi}
\affiliation{Department of Physics, Sharif University of Technology, Tehran, P.O.Box: 11365-9161, Iran}
\author{M.~A.~Rajabpour\footnote{e-mail: rajabpour@sissa.it}}
\affiliation{SISSA and INFN, \textit{Sezione di Trieste},  via Bonomea 265, 34136 Trieste, Italy}

\begin{abstract}
We study the fractal properties of the $2d$ discrete scale invariant (DSI) rough surfaces. The contour lines of these rough surfaces show clear DSI. In the appropriate limit the DSI surfaces converge to the scale invariant rough surfaces. The fractal properties of the $2d$ DSI rough surfaces apart from possessing the discrete scale invariance property follow the properties of the contour lines of the corresponding scale invariant rough surfaces. We check this hypothesis by calculating numerous fractal exponents of the contour lines by using numerical calculations. Apart from calculating the known scaling exponents some other new fractal exponents are also  calculated.
\end{abstract}

\maketitle

\section{Introduction}\
 Discrete scale invariance (DSI) is a ubiquitous phenomenon in human
  made fractals  and also in the fractals of nature.
The list of applications of discrete scale invariance covers phenomena
 in a wide range of topics such as  diffusion in anisotropic quenched random lattices \cite{Staufer},
growth processes and rupture \cite{AFSS}, non-extensive statistical
physics \cite{non-extensive}, econophysics \cite{econo1,econo2}, sociophysics \cite{ZHSD}, as well as quenched disordered systems \cite{SS} and cosmic lacunarity \cite{cosmic}, for  review of
the concept and other applications see \cite{sornette1,sornette2,DGS}.
 The signature of the presence of   DSI is complex exponent which manifest itself in log-periodic
corrections to scaling. In some cases the log-periodic oscillations appear in
the time dependence of some physical quantities such as the energy release on the approach of impending rupture \cite{ALSS} and earthquakes \cite {SSS} and in some other cases the DSI appears in the
geometry of the system,
the most famous cases being animals \cite{SS}, diffusion limited aggregation (DLA) \cite{SJAMS} and sandpile model on Sierpinski gasket \cite{sandpile}. Recently we found an other example in surface science, some of the rough surfaces as well as their contour lines show clear DSI in their geometry \cite{GR}. The rough surfaces have been the subject of intense studies for many years \cite{BS}.  The first predictions of the scaling exponents of the contour lines of the scale invariant rough surfaces appeared in \cite{KH}. The scaling exponents predicted in \cite{KH} were confirmed numerically in \cite{KHS} and the different applications of the findings in glassy interfaces and turbulence were discussed in \cite{ZKMM}. After the invention of the stochastic Schramm-Loewner evolution \cite{Schramm} which describes scale invariant curves in two dimensions a revival appeared in the field and many new properties of the contour lines were investigated, for example see \cite{SRR,RV,SDR}. One of the most important findings of the paper \cite{KH} is the dependence of the different exponents of the contour lines to the only universal parameter in the system, i.e. roughness (\textit{Hurst}) exponent.  They showed that for the full scale invariant rough surfaces the fractal dimension of all of the contours $d$, the fractal dimension of one contour $D_{f}$ and the length distribution exponent $\tau$ of the contour lines are just simple functions of the roughness exponent. Altough there is no theoretical proof for the scaling relations found in \cite{KH} they were confirmed in many different numerical studies \cite{KHS, RV}.\\
The building block of the functions with DSI is Weierstrass-Mandelbrot (WM) function which was studied by Berry and Lewis in \cite{BL}. It is a random continuous non differentiable mono fractal function that one can see it  in a wide range of physical phenomena such as sediment, turbulence \cite{sornette1,sornette2} and propagation and localization of waves in fractal media \cite{Localiz}. The function is scale invariant just for particular value of the scale parameter $\gamma$. It was shown in \cite{BL,GR} that in the $\gamma \to 1$ limit the WM function converges to the Brownian motion. In the other words one can think about WM function as the perturbed Brownian motion which the perturbing parameter is the $\gamma$. The same argument is also true  for the 2d WM function \cite{AB, Falconer} and one can think about the 2d WM function as the  mono fractal rough surface with the DSI property for the specific value of $\gamma$. It was shown in \cite{GR} that up to the fractal dimension calculations many properties of the WM function is the same as the Brownian motion counterpart. What we are going to do in this paper is confirming this idea for the contour lines of the 2d WM function. We will show numerically that many exponents of the contour lines of the 2d WM function are the same as the exponents of the contour lines of the corresponding 2d Brownian sheet. In other words we confirm the same scaling relations as the paper \cite{KH} for our discrete mono scale invariant contour lines.\\
The structure of the paper is as follows: In the next section we will introduce 1d and 2d Weierstrass-Mandelbrot function and we will discuss the DSI property of the surfaces. The third section is basically for fixing the notation and introducing different exponents, most of the materials of this section were already discussed in \cite{KH, KHS}. In the forth section we will numerically confirm all the proposed scaling exponents and relations of the third section  and we will show the universality of those relations. In the last section we will summarize our findings and we will give some comments about the possible future directions.

\section{Definition of 1d and 2d Weierstrass-Mandelbrot function}\

The Weierstrass-Mendelbrot (WM) function is a continuous non differentiable function. This function has discrete scale invariance property which is a weaker kind of scale invariance. In other words for the measurable quantity $\mathcal O$; the relation ${\mathcal O}(\lambda x)=\mu(\lambda){\mathcal O}(x)$ occurs only for a special value of $\lambda $. The definition of the one dimensional WM function is
\begin{eqnarray}\label{general_one_WM}
W(t)=\sum^{\infty }_{n=-\infty }{\frac{{g(\gamma ^n t)}e^{i\phi_n}}{{\gamma }^{\left(2-D\right)n}}};
\end{eqnarray}
where $g(x)$ is a generic periodic function, $\gamma>1$ , $1<D<2$ and $\phi_n$ is an arbitrary phase \cite{BL}. This function does not change under scaling transformation $t\to bt$ if and only if $b=\gamma $. In this case WM function has self-affine scaling law
\begin{eqnarray}\label{WM_scaling}
W(\gamma t)={\gamma }^{2-D}W(t)=\gamma^{H }W(t),
\end{eqnarray}
where the parameter $H =2-D$ is the \textit{Hurst} exponent. The self affine scaling property of the WM function can be also seen in the correlation function of the increment $\chi=\Delta W(t,\tau)=W(t+\tau)-W(t)$ of the process which is
\begin{eqnarray}\label{WM_correl}
 C(\tau)={\left\langle {\left|\chi\right|}^2\right\rangle }_\phi = \sum^{\infty }_{n=-\infty }{\frac{\vert g\left(\gamma^n(t+\tau)\right)-g\left(\gamma^n(t)\right)\vert ^2}{{\gamma }^{2\left(2-D\right)n}}},
\end{eqnarray}
where phase $\phi_n$ picked up from the uniform random distributed numbers. Since the condition $C(\gamma \tau)={\gamma }^{2(2-D)}C(\tau )$ is true for all periodic functions $g(x)$, therefore the correlation function is also scale invariant.

The Weierstrass-Mandelbrot function was generalized to two dimensions in \cite{AB} as follows
\begin{eqnarray}\label{Two_WM}
W(x,y)=C\sum_{\begin{array}{cc}
m\in[1,M] \\n\in(-\infty,\infty) \end{array}}
A_mB_nG(r\eta_{m,n} ) e^{ i\phi_{m,n}};
\end{eqnarray} 
where $C=\left({\text{ln}\gamma}/{M}\right)^{1/2}$, $B_n={\left(k_0{\gamma }^n\right)}^{D-3}$, $\eta_{m,n}=k_0\cos(\theta -{\alpha }_m){\gamma}^n$, $r=\left(x^2+y^2\right)^{1/2}$ and $\theta=\tan^{-1}\left({y}/{x}\right)$. The normalization factor $C$ is chosen so that the correlation function $V\left({\mathbf \rho }\right)$ be finite in the limits $M\to \infty $ and $\gamma \to 1$. The $A_m$ factor can control the isotropy of WM surface. We have an isotropic rough surface when $A_m=A$. The wave number $k_0=\frac{2\pi }{L}$ corresponds to system size $L$ and for each $m$ indices there is one phase ${\alpha }_m$ that may be deterministic or stochastic. In this study ${\alpha }_m=\frac{\pi m}{M}$ and $\phi_{m,n}$ is selected randomly from the interval $[0,2\pi ]$ with uniform distribution. 

In numerical methods the two parameters $M$ and $n$ should be finite but large enough that $W\left(x,y\right)$  does not change with increasing these parameters. Fractal dimension $D$ and \textit{Hurst} exponent $H =3-D$ are two important parameters in simulating WM function, for an example of 2d WM function see Fig. ~1. The surfaces with larger $H $ seems smoother than the surfaces with smaller $H $.  

\begin{figure} [htb]
\centering
\includegraphics[width=0.4\textwidth]{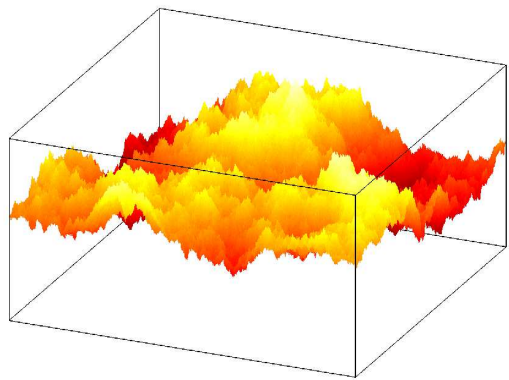}
\caption{(Color online) WM rough surfaces with $H=0.4$ and $\gamma=1.5$. Light yellow (light Gray) regions correspond to larger heights and dark red (dark Gray) regions correspond to smaller heights.}
\label{Figure:1}
\end{figure}

The correlation function of the two dimensional WM function $C(\mathbf{\rho})={\left\langle {\left|   W(\mathbf{r}+\mathbf{\rho})-W(\mathbf{r})  \right|}^2\right\rangle }_\phi$
obeys the scaling law $C(\gamma {\mathbf \rho })={\gamma }^{2(3-D)}C({\mathbf \rho })$ where $D$ is the fractal dimension and $2<D<3$. This shows that $C({\mathbf \rho })$ is also a self similar function.

One of the capabilities of WM function is that some of the statistical properties of this function do not have any sensitivity to the form of $G(x)$. We use $1-e^{ix}$ and $\sin^{-1}(\sin(x))$ as $G(x)$ in Eq. (\ref{Two_WM}) and we will show that the statistical properties of the contour lines of 2d WM function depend only to the \textit{Hurst} exponent of this process.


\section{Scaling laws of the contour lines  }\

For a given stochastic scale invariant rough surface with the height $h(\mathbf{x})$, a level set $h(\mathbf{x})=h_0$ for different values of $h_0$ consists of many closed non-intersecting loops. These loops are scale invariant and their size distribution is characterized by a few scaling functions and scaling exponents. For example the contour line properties can be described by the loop correlation function $G_l(\mathbf{r})$.  The loop correlation function measures the probability that two points  separated with distance $\mathbf{r}$ in the plane lie on the same contour. Rotational invariance of the contour lines forces  $G_l(r)$ to depend only on $\vert \mathbf{r} \vert$. This function for the contours on the lattice with grid size $a$ and in the limit $r>>a$ has the scaling behavior
\begin{eqnarray}\label{G_scaling}
G_l(r) \sim \frac{1}{r^{2x_l}},
\end{eqnarray}
where $x_l$ is the loop correlation exponent.

Another measure associated with a given contour loop ensemble is $G_s(r)$.  
This probability is called the two point correlation function for contour lines with length $s$. Scaling properties of the contours force $G_s(r)$ to scale with $s$ and $r$ as
\begin{eqnarray}\label{G_s_scaling}
G_s(r) \sim s^m \vert r \vert ^{-n} f_{G_S}\left( r/R \right), 
\end{eqnarray}
where $m$ and $n$ are two unknown exponents and $f_{G_S}\left( r/R \right)$ is a scaling function. In this function we must scale $r$ by the typical diameter  $R$  which is called radius of gyration. For one loop with $N$ discrete points $\left\lbrace (x_1,y_1) ,..., (x_N,y_N)\right\rbrace $, the radius of gyration  $R$  is defined by
\begin{eqnarray}\label{Gyration_radius}
R^2=\frac{1}{N} \sum_{i=1}^N \Big{(}(x_i-x_c)^2+(y_i-y_c)^2\Big{)},
\end{eqnarray}
where $x_c=\frac{1}{N}\sum_{i=1}^N x_i$ and $y_c=\frac{1}{N}\sum_{i=1}^N y_i$ are the central mass coordinates.

The 2d WM function is a self affine surface. A key consequence of this result is that, the contour lines with perimeter $s$ and radius $R$ of such surfaces are self-similar. When these lines are scale invariant one can determine the fractal dimension as the exponent in the perimeter-radius relation. The relation between contour length $s$ and its radius $R$ is

\begin{eqnarray}\label{scaling_of_s}
s \sim R^{D_f},
\end{eqnarray}
where $D_f$ is the fractal dimension, $R$ is defined by Eq. (\ref{Gyration_radius}) and $s$ is measured with a ruler of length $a$. It is a right time to mention that the $D_{f}$ is the fractal dimension of one contour and it is different from the fractal dimension of  all the level set $d$. The latter one can be derived from the fractal dimention of the rough surface which is $3-H$ and the intersection rule of the Mandelbrot. Since  the level set, the set of the contour lines, is the intersection of the rough surface with a two dimensional surface one can derive the following relation 
\begin{eqnarray}\label{level set fractal dimension}
d=2-H.
\end{eqnarray}

For a given self similar loop ensemble one can define the probability distribution of contour lengths $\tilde{P}(s)$. This function is a measure for the loops with length $s$ and follows the power law
\begin{eqnarray}\label{scaling_of_P}
\tilde{P}(s) \sim s^{-\tau},
\end{eqnarray}
where $\tau$ is a scaling exponent. The above power law functions (Eqs. (\ref{G_s_scaling}), (\ref{scaling_of_s}) and (\ref{scaling_of_P})), introduce the geometrical exponents $x_l$, $D_f$ and $\tau$. These exponents depend only on the roughness exponent $H$ \cite{KH}.  

One can derive mean loop length $<s>$ and the probability distribution of length $\tilde{P}(s)$ from another quantity $\tilde{n}(R,s)$. This function is the probability distribution of contours with length $(s,s+ds)$ and radius $(R,R+dR)$. Scale invariance of the contours forces $\tilde{n}(R,s)$ to scale with $s$ and $R$ as
\begin{eqnarray}\label{scaling_n}
\tilde{n}(R,s) \sim s^{-y}f_n \left(s/R^{D_f}\right).
\end{eqnarray}
In the above relation $y$ is related to the fractal dimension $D_f$ and the length distribution exponent $\tau$. It is obvious that
\begin{eqnarray}\label{mean_length_from_n}\label{P_from_n}
<s>(R) &=&\frac{\int^{\infty }_0{ds \tilde{n}(R,s)s}}{\int^{\infty }_0{ds\tilde{n}(R,s)}},
\\ \tilde{P}(s)&=&\int^{\infty }_0{dR\tilde{n}(R,s)}.
\end{eqnarray}
The above equations and Eq. (\ref{scaling_n}) yield $\tilde{P}(s)\sim s^{-y+1/D_f}$, which we can equate this with Eq. (\ref{scaling_of_P}), and obtain $y=\tau+1/D_f$. Notice that, one can derive scaling law (\ref{scaling_of_s}) from Eqs. (\ref{scaling_n}) and (\ref{mean_length_from_n}).

Following \cite{KH} one can determine the relation between all geometrical exponents. In order to find the first scaling  relation, one can use from the scale invariance of the rough surfaces, i.e. $h(b\mathbf{x})=b^{H}h(\mathbf{x})$ under $\mathbf{x} \rightarrow b\mathbf{x}$,  and invariance of $\tilde{n}(R)$ under rescaling, i.e. $\tilde{n}(R/b)=b^{3-H}\tilde{n}(R)$, then
\begin{eqnarray}\label{scaling_nR_new}
\tilde{n}(R) \sim R^{-3+H}.
\end{eqnarray}
On the other hand, the number of loops with radius in $(R,R+dR)$ can be derived from $\tilde{n}(R)=\int^{\infty}_0ds\tilde{n}(R,s)$, where it obeys the scaling law
\begin{eqnarray}\label{scaling_nR}
\tilde{n}(R) \sim R^{-(1+D_f(\tau-1))}.
\end{eqnarray}
Equations  (\ref{scaling_nR_new}) and (\ref{scaling_nR}) lead to the first scaling relation
\begin{eqnarray}\label{hyperscaling}
D_f(\tau -1)=2-H.
\end{eqnarray}
This relation is called hyperscaling relation; it is derived for the scale invariant rough surfaces. The above argument is also true for the discrete scale invariant surfaces such as 2d WM functions. For the DSI rough surfaces such as Eq. (\ref{Two_WM}), condition $h(b\mathbf{x})=b^{H}h(\mathbf{x})$ is true just for a special value of $b=\gamma$. 

For a self affine random field, there is another scaling relation that connect the three geometrical exponents $D_f$, $\tau$ and $x_l$. This relation is derived from a sum rule \cite{KH}. Integrating (\ref{G_s_scaling}) over all lattice points gives the total number of points on the loop
\begin{eqnarray}\label{s}
s=\int d^2\mathbf{r}G_s(\mathbf{r}) \sim s^{(2-n)/D_f+m}.
\end{eqnarray}
The above equality gives one relation between the exponents $n$ and $m$ as
\begin{eqnarray}\label{relation_n_m}
2-n=D_f(1-m).
\end{eqnarray} 
The integration of $G_s(\mathbf{r})$ over loop probability distribution function $P(s)=s\tilde{P}(s)$ gives the total loop correlation function $G_l(r)$ 
\begin{eqnarray}\label{G_l}
G_l(r)=\int dsP(s)G_s(\mathbf{r}) \sim r^{-n} \left( r^{D_f} \right)^{m+2-\tau},
\end{eqnarray}
where after using Eqs. (\ref{G_scaling}) and (\ref{G_l}) gives another relation between $m$ and $n$ as
\begin{eqnarray}\label{relation_n_m_new}
-n+D_f(m+2-\tau)=-2x_l.
\end{eqnarray}
Elimination of $m$ and $n$ from equations (\ref{relation_n_m}) and (\ref{relation_n_m_new}) leads to the second scaling relation
\begin{eqnarray}\label{sum_rule}
D_f(3-\tau)=2-2x_l.
\end{eqnarray} 
If we combine the two scaling relations (\ref{hyperscaling}) and (\ref{sum_rule}) we can find two exponents $D_f$ and $\tau$ as a function of the \textit{Hurst} exponent $H$ and the loop correlation exponent $x_l$,
\begin{eqnarray}\label{D}
D_f=2-x_l-\frac{H}{2},
\end{eqnarray}
and
\begin{eqnarray}\label{tau}
\tau =1+\frac{2-H}{2-x_l-H/2}.
\end{eqnarray}

In the next section we will numerically calculate all the introduced exponents related to the contour lines of DSI rough surfaces and we will show that the two scaling relations (\ref{hyperscaling}) and (\ref{sum_rule}) are valid for generic 2d-WM function.

\section{Numerical results}\

To simulate the 2d WM function on the discrete lattice we used  lattices on square grid with sizes 1024 and 2048. In equation (\ref{Two_WM}) we set $M=20$ and $-n_{min}=n_{max}\ =50$. After setting $\gamma {\rm =1.5}$ we simulated the WM rough surfaces with the  \textit{Hurst} parameter in the range of  $H=\{0.0,0.1,0.2,0.3,0.4,0.5,0.6\}$ for two periodic functions $G(x)=(1-e^{ix})$ and $G(x)=\sin^{-1}(\sin(x))$ . In each case, all numerical tests have been done using from 250 realizations.
The contouring algorithm followed from \cite{RV} to generate contour lines of rough surfaces. In Fig.~2 one can see that the number of loops increases by decreasing roughness exponent. 

\begin{figure*}[htb]
\begin{center}
\includegraphics[angle=0,scale=0.45]{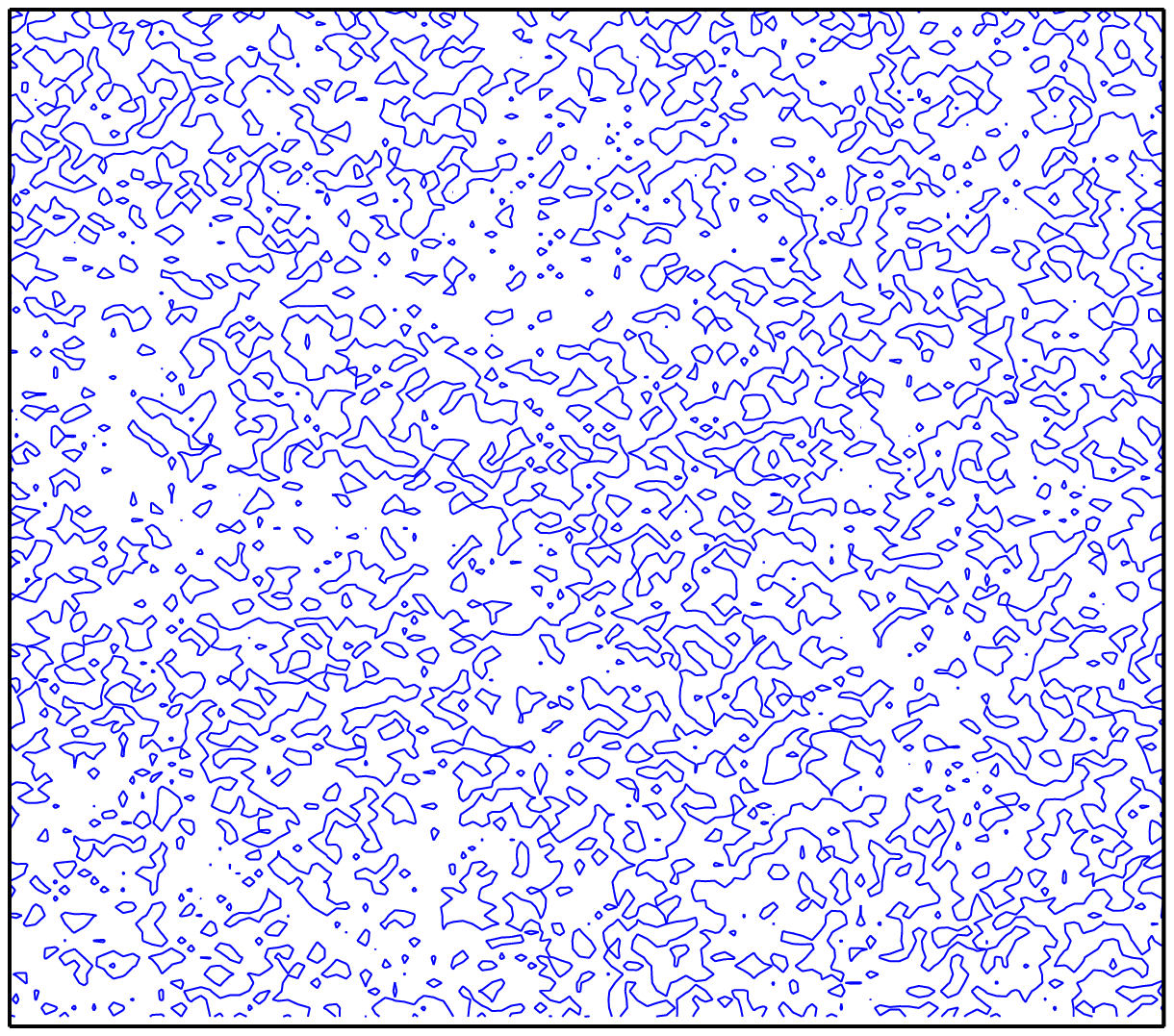}
\includegraphics[angle=0,scale=0.45]{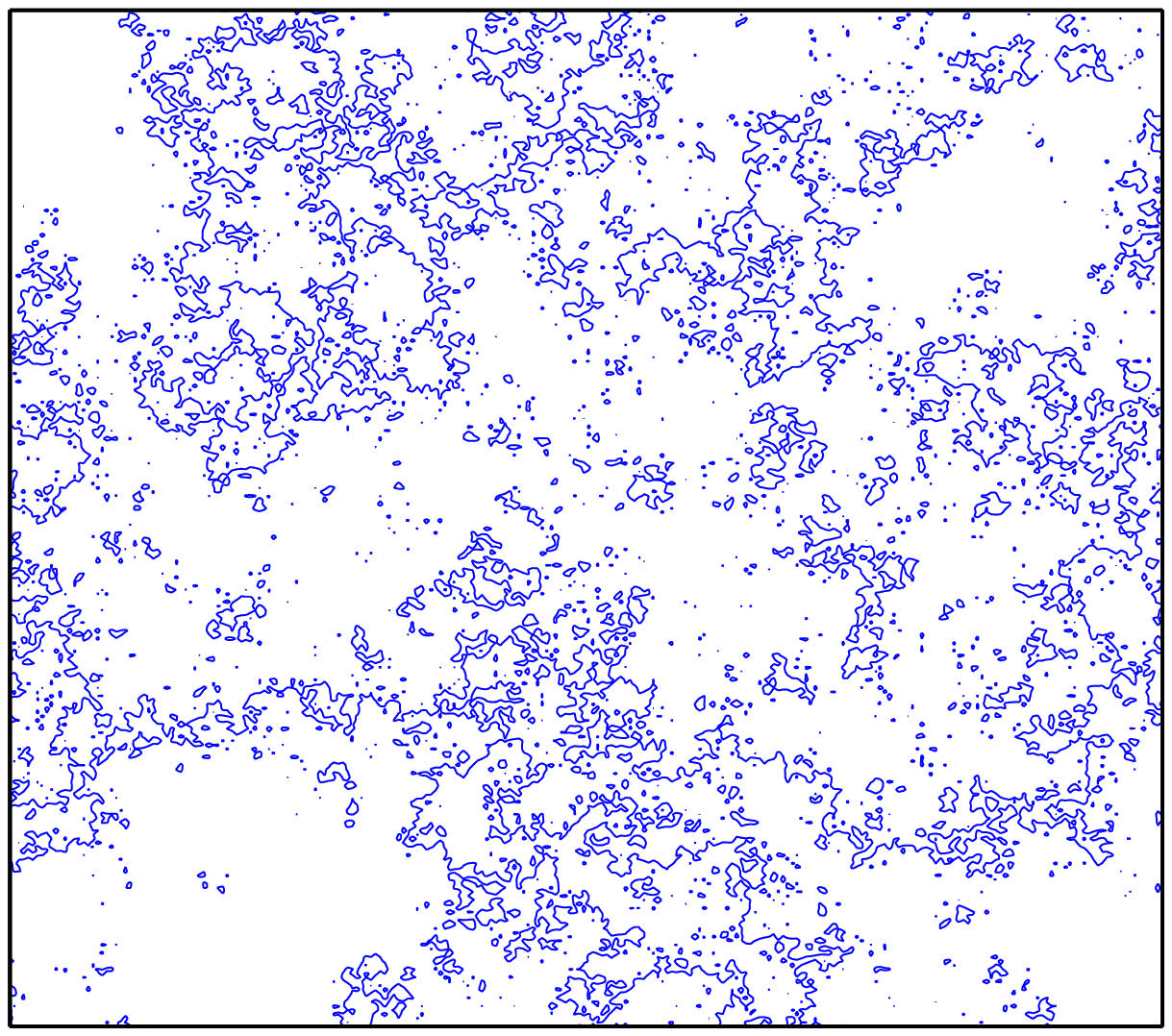}
\includegraphics[angle=0,scale=0.45]{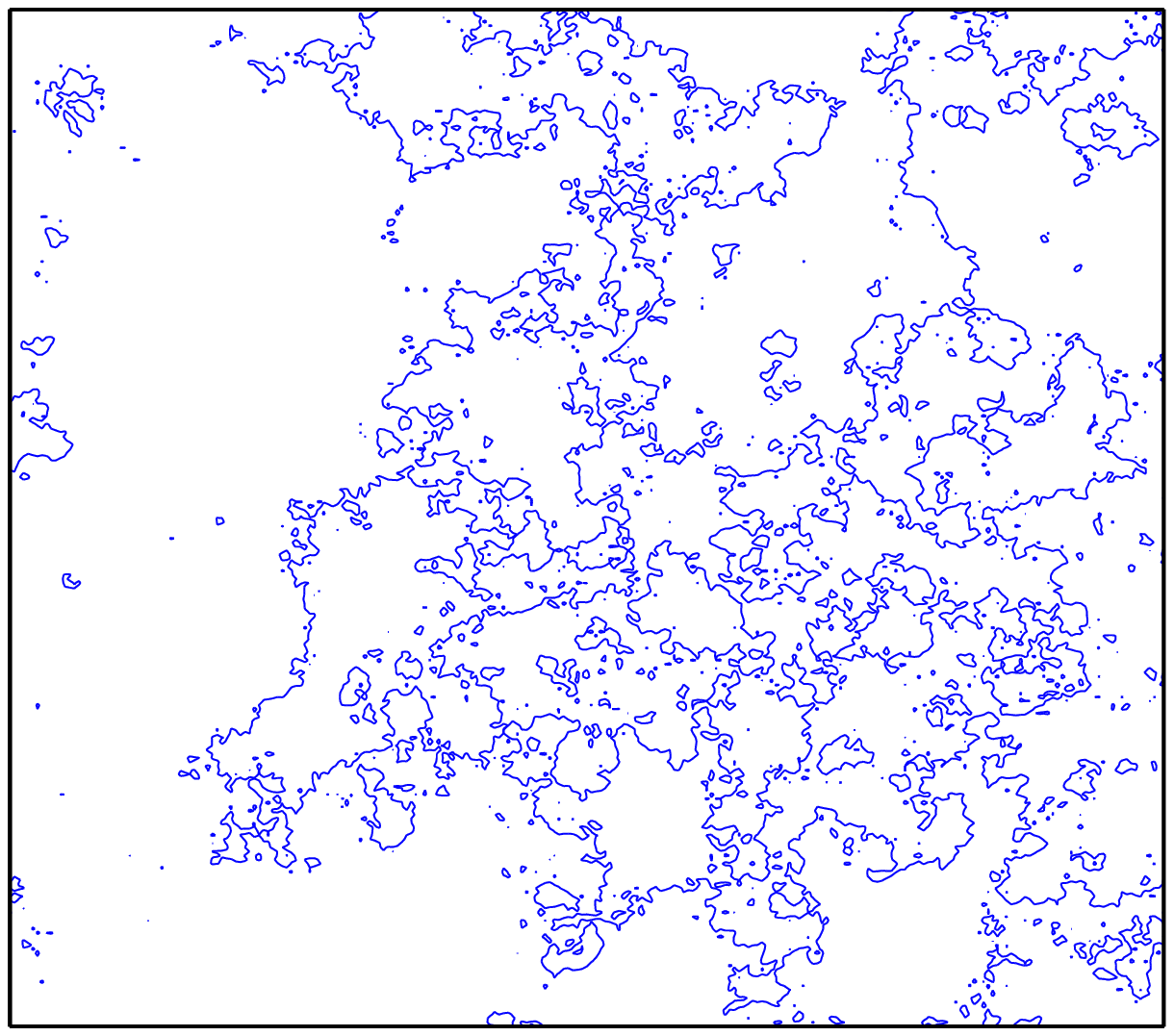}
\includegraphics[angle=0,scale=0.45]{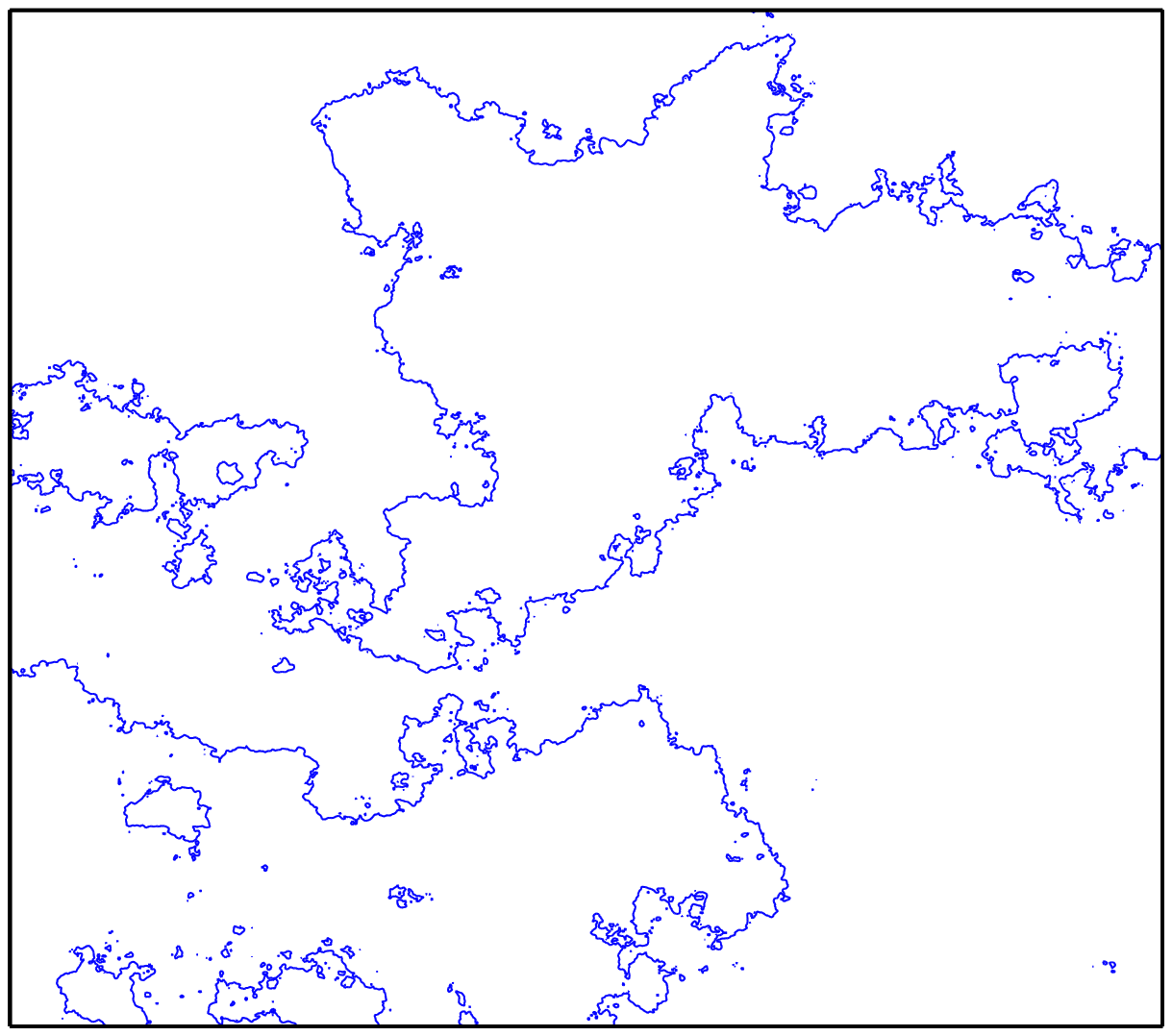}
\caption{(Color online) Small parts of contour lines of 2d WM rough surface with $G(x)=(1-e^{ix})$ and lattice size $1024^2$ . Top left ($H=0.0$), right ($H=0.2$), and bottom left
($H=0.4$), right ($H=0.6$).}
\end{center}
\label{Figure:2}
\end{figure*}

We would like to measure the exponents $d$, $D_f$, $\tau$ and $x_l$ with a large enough loop ensemble for the mentioned  values of the \textit{Hurst} exponents. Before starting numerical computation of the geometrical exponents, it is important to see the  DSI property in the contour lines of such discrete scale invariant rough surfaces. At the beginning we will show that the contour lines of DSI rough surfaces are discrete scale invariant. Existence of DSI in the loop ensemble can be studied by the Lomb normalized priodogram test.

\subsection{The Lomb normalized periodogram}\
The 2d Weierstrass Mandelbrot function (\ref{Two_WM}) is a discrete scale invariant function which can be constructed by using periodic function $G(x)$ with infinite number of discrete frequencies $\omega_n=\gamma^n$. Each frequency contributes in (\ref{Two_WM}) with weight $\gamma^{-Hn}$ . To detect these frequencies one can use the Lomb normalized periodogram.

For a given time series  $X_i\equiv X(t_i)$ with $N$ data points, this kind of spectral analysis fitting these data points to a sine function of varying frequencies by least square method \cite{lomb}. Lomb analysis measures spectral power $P(\omega)$ as a function of angular frequency $\omega$ as
 \begin{eqnarray}\label{lomb}
P_N(\omega) = \frac{1}{2\sigma^2}\Big(  \frac {[\sum_j(X_j-\overline{X})\cos \omega (t_j-\delta)]^2}{\sum_j\cos^2\omega(t_j-\delta)} +\nonumber\\ \frac{[\sum_j(X_j-\overline{X})\sin\omega(t_j-\delta)]^2}{\sum_j\sin^2\omega(t_j-\delta)}\Big);
\end{eqnarray}
where $\overline{X}=\frac{1}{N}\sum^N_1X_i$, $\sigma^2=\frac{1}{N-1}\sum^N_1(X_i-\overline{X})^2$ and $\delta$ is defined by the relation $\tan(2 \omega \delta) = \frac{\sum_j \sin 2 \omega t_j}{\sum_j \cos 2 \omega t_j}$.  

We used from Lomb test to show that the contour lines of the DSI rough surfaces are discrete scale invariant. For a closed loop (see Fig. ~3), with $N$ data points $\lbrace(x_1,y_1),...,(x_N,y_N)\rbrace$, the distance from the center of mass; i.e. $r(i)=\Big{(}(x_i-x_c)^2+(y_i-y_c)^2\Big{)}^{1/2}$, where $x_c=\frac{1}{N}\sum_{i=1}^{N}x_i$ and $y_c=\frac{1}{N}\sum_{i=1}^{N}y_i$ are the center of mass coordinates; is a good quantity to detect the discrete frequencies.
Using Lomb test we measured $\gamma_{exp}=\langle\omega_{i+1}/\omega_i\rangle=1.55 \pm 0.04$ where $\omega_{i+1}$ and $\omega_{i}$ are two frequencies that $P(\omega)$ has sharp picks, results are shown in Fig. ~4. This result is in a good agreement with the theoretical value of $\gamma=1.5$.

\begin{figure}[htb]
    \centering
        \includegraphics[width=0.5\textwidth]{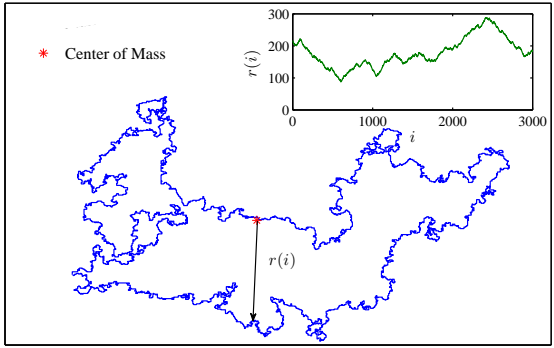}
\caption{(Color online) Largest contour line of 2d WM rough surfaces with $N=6\times10^3$, $D=2.5$ and $\gamma=1.5$. Distance $r(i)$ for each point $i$ from the center of mass is shown in the inset.} 
\label{Figure:3}
\end{figure}

\begin{figure}[htb]
    \centering
        \includegraphics[width=0.5\textwidth]{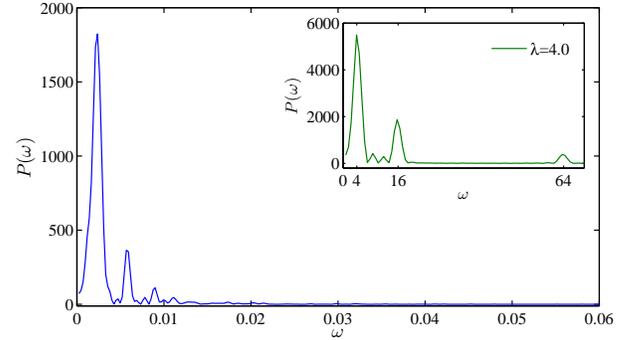}
\caption{(Color online) Lomb test on the contour line of 2D WM surfaces with $D=2.5$ and $\gamma=1.5$. Lomb test on 1d WM time series with $\gamma=4.0$, is shown in the inset. Sharp picks show that the contour lines of WM surfaces are DSI.}
 \label{Figure:4}
\end{figure}

\subsection{Loop correlation exponents} \
In this subsection we will numerically find the most fundamental exponents of the contour lines of the rough surfaces, i.e. $x_l$ and $n$, by analyzing the functions $G_s(r)$ and $G_{l}(r)$.

\subsubsection{Scaling exponents in $G_{l}(r)$}\
To find the correlation function from a given loop ensemble we follow the algorithm described in \cite{KH}. Using scaling properties of loop correlation function we can measure correlation exponent $x_l$. For the large class of rough surfaces $x_l=\frac{1}{2}$, this scaling exponent is independent of $H$ \cite{KH}.  Our numerical tests show that $x_l=0.5 \pm 0.05$ for all the WM rough surfaces and it is not a function of $H$ or $G(x)$ as shown in Fig.~5.

\begin{figure}[htb]
    \centering
        \includegraphics[width=0.5\textwidth]{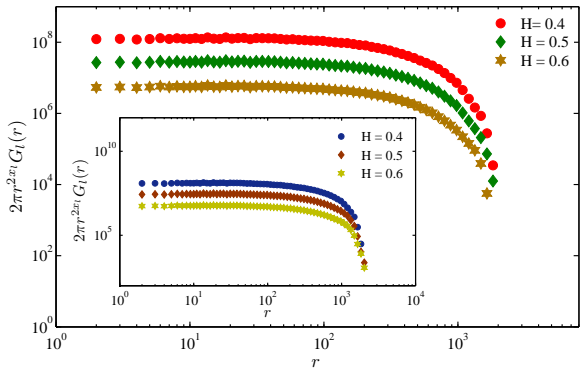}
\caption{(Color online) The loop correlation function for the DSI rough surfaces with $H=0.4$, $0.5$ and $0.6$ (from  top to bottom) for $G(x)=1-e^{ix}$. The loop correlation function for $G(x)=\sin^{-1}(\sin(x))$ is shown in the inset.  The plots were derived for the system size $L=2048$ and $\gamma=1.5$.} 
\label{Figure:5}
\end{figure}

\subsubsection{Scaling exponents in $G_s(r)$}\
The scaling behavior of the two point correlation function of the contours with length $s$ is defined in Eq. (\ref{G_s_scaling}). The exponent $n$ can be measured from the power law dependence of $G_s(r)$ with respect to $r$ for a fixed value of the perimeter $s$, see Fig.~6. For this purpose the contour loops with the perimeter in the range $(s,\alpha s)$ ($\alpha>1$) are considered and  $G_s(r)$ is measured as a function of $r$ for the different values of $s$, see Table ~1. There is neither theoretical prediction for this exponent and nor any numerical estimates. Our numerical calculation shows that the exponent $n$ closely follows the relation $\frac{1+H}{2}$, see ~Fig.~7 which means that the exponent $m$ is very small compared to the exponent $n$. Our numerical calculation shows that the same values can also be derived for the scaling mono fractal rough surfaces.

\begin{figure}[htb]
    \centering
        \includegraphics[width=0.5\textwidth]{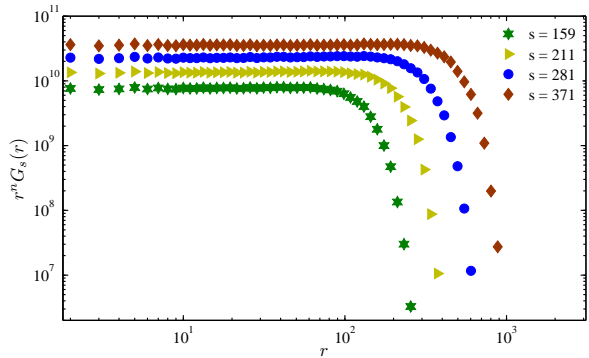}
\caption{(Color online) The scaling behaviour of  $G_s(r)$ as a function of $r$ in the DSI rough surfaces for $G(x)=1-e^{ix}$  with  $L=2048$, $H=0.5$ and $\gamma=1.5$.}
\label{Figure:6}
\end{figure}

\begin{figure}[htb]
    \centering
        \includegraphics[width=0.5\textwidth]{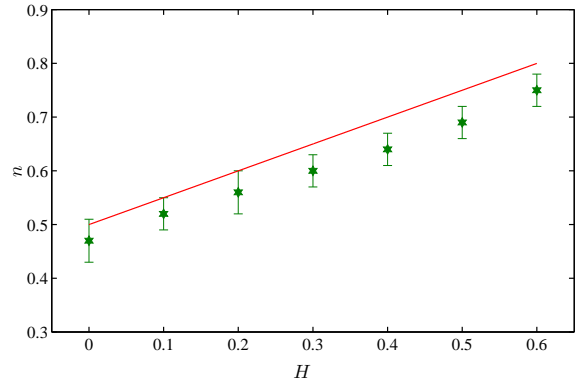}
\caption{(Color online) The exponent $n$ with respect to the roughness exponent $H$ for $G(x)=1-e^{ix}$. The slop of the best fit line is $0.45\pm0.03$. The smooth line is the line $n=\frac{1+H}{2}$.}
\label{Figure:7}
\end{figure}

\begin{table}[htp]
\begin{center}
\begin{tabular}{|c|c|c|c|}\hline
$H$ &  $G(x)=1-e^{ix}$  & $G(x)=\sin^{-1}(\sin(x))$ \\
\hline
$0.0$  &$0.46 \pm {0.04}$ &$0.47 \pm {0.03}$    \\
\hline
$0.1$  &$0.53 \pm {0.03}$ & $0.52 \pm {0.03}$
\\ \hline
$0.2$  &$0.56 \pm {0.04}$ & $0.56 \pm {0.02}$
\\ \hline
$0.3$  &$0.59 \pm {0.03}$ & $0.59 \pm {0.04}$
\\ \hline
$0.4$  &$0.64 \pm {0.03}$ & $0.64 \pm {0.03}$\\
\hline
$0.5$  &$0.69 \pm {0.03}$ & $0.69 \pm {0.04}$\\
\hline
$0.6$  &$0.74 \pm {0.03}$ & $0.75 \pm {0.03}$\\

\hline
\end{tabular}
\end{center}
\caption{\label{tab1} The numerical values of the scaling exponent $n$ 
derived from the  two point correlation function $G_s(r)$.}
\end{table}

\subsection{Fractal dimensions}\
To study the fractal properties of contour lines, the fractal dimension of loops $D_f$ and the fractal dimension of all the contours $d$, in DSI rough surfaces one can use from three related approaches. In the first method one can use from self-similar properties of contours to measure fractal dimensions. The second method is using cumulative distribution of area and the last one is the Zipf's law. We used these three methods  to find the two geometrical exponents $D_f$ and $d$.

\begin{enumerate}
\item[\textit{a. }]\textit{Length-radius scaling relation: }
\end{enumerate}

In order to find the fractal dimension of the contours $D_f$ we used directly from the scaling relation between loop perimeter $s$ and the radius of gyration $R$ according to Eq. (\ref{scaling_of_s}). Scaling behavior of the averaged loop length $<s>$ with respect to $R$ in $\text{log-log}$ plot is shown in Fig.~8. The geometrical exponent $D_f$ is measured using the least-square linear fit in the scaling regime ($10<R<100$).

\begin{figure}[htb]
    \centering
        \includegraphics[width=0.5\textwidth]{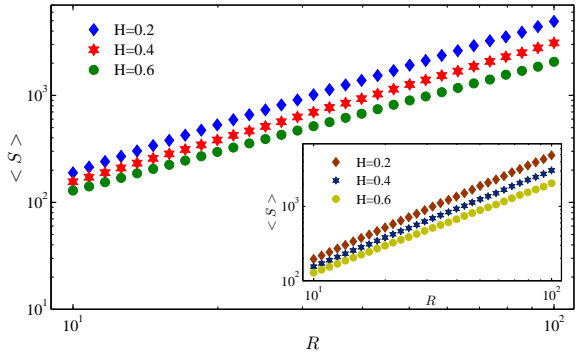}
\caption{(Color online) The scaling relation between the averaged loop length $<s>$ and the radius of gyration $R$ for contour lines of the DSI rough surfaces with $G(x)=1-e^{ix}$. The same scaling behavior for $G(x)=\sin^{-1}(\sin(x))$  is shown in the inset.}
\label{Figure:8}
\end{figure}

The fractal dimension of all contours $d$ can be found by box counting method \cite{Mandel2}. This method is based on the self similar properties of all contours in the level set. For a self similar object there is a scaling law between the number of boxes $N(L)$  with size $L$ needed to cover the fractal object
 \begin{eqnarray}\label{Boxcounting}
N(L) \sim L^{-d}. 
\end{eqnarray}
As shown in Fig.~9, we find excellent agreement between a measured values of $D_f$ and $d$ with theoretical predictions.  

\begin{figure}[htb]
    \centering
        \includegraphics[width=0.5\textwidth]{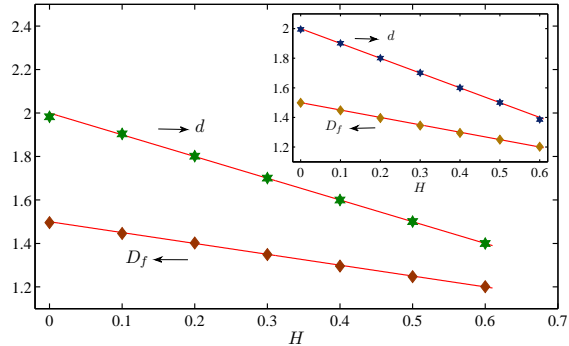}
\caption{(Color online) The fractal dimension of all contours $d$ and  the geometrical exponent $D_f$ as a function of $H$ for the  DSI surfaces with $G(x)=1-e^{ix}$. These scaling exponents for  $G(x)=\sin^{-1}(\sin(x))$ are shown in the inset. The solid lines correspond to the theoretical predictions. }
\label{Figure:9}
\end{figure}

\begin{enumerate}
\item[\textit{b. }]\textit{Cumulative distribution of areas: }
\end{enumerate}

The cumulative distribution of the loop area is a good measure to describe the fractal dimension of the set of all contour lines $d$ \cite{RV}. For a self affine random field the number of contours with area greater than $A$  has the scaling form
\begin{eqnarray}\label{cumulative}
N_>(A)\sim A^{-d/2},
\end{eqnarray}
where $d=2-H$.  An example of the scaling behaviour in cumulative distribution of area is shown in Fig.~10. Here, we plot $N_>(A)$ as a function of $A$ for $H=0.4$ in $\text{log-log}$ plot, to show how they follow Eq. (\ref{cumulative}). In Table \ref{tab2} we report the scaling exponent $d/2$ for different values of $H$.

\begin{figure}[htb]
    \centering
        \includegraphics[width=0.5\textwidth]{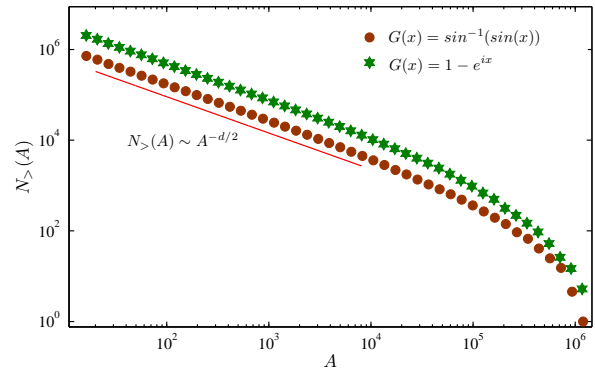}
\caption{(Color online) Cumulative number of loops whose area is greater than $A$ for the DSI rough surfaces with $G(x)=1-e^{ix}$ and $G(x)=\sin^{-1}(\sin(x))$  with $H=0.4$, system size $L=2048$ and $\gamma=1.5$. } 
\label{Figure:10}
\end{figure}

\begin{table}[htp]
\begin{center}
\begin{tabular}{|c|c|c|c|}\hline
$H$& Theory &  $G(x)=1-e^{ix}$  & $G(x)=\sin^{-1}(\sin(x))$ \\
\hline
$0.0$ & 1.00 &$1.00 \pm {0.01}$ &$1.00 \pm {0.01}$    \\
\hline
$0.1$ & 0.95 &$0.96 \pm {0.01}$ & $0.96 \pm {0.01}$
\\ \hline
$0.2$ & 0.90 &$0.91 \pm {0.01}$ & $0.90 \pm {0.01}$
\\ \hline
$0.3$ & 0.85 &$0.86 \pm {0.01}$ & $0.86 \pm {0.01}$
\\ \hline
$0.4$ & 0.80 &$0.81 \pm {0.01}$ & $0.81 \pm {0.01}$\\
\hline
$0.5$ & 0.75 &$0.76 \pm {0.01}$ & $0.77 \pm {0.02}$\\
\hline
$0.6$ & 0.70 &$0.71 \pm {0.01}$ & $0.71 \pm {0.01}$\\

\hline
\end{tabular}
\end{center}
\caption{\label{tab2} The best fit values of  the
exponent $d/2$
derived from the scaling laws of cumulative distribution $N_>(A)$. }
\end{table}

\begin{enumerate}
\item[\textit{c. }]\textit{Zipf's laws }
\end{enumerate}

\begin{figure}[htb]
\begin{center}
\includegraphics[angle=0,scale=0.35]{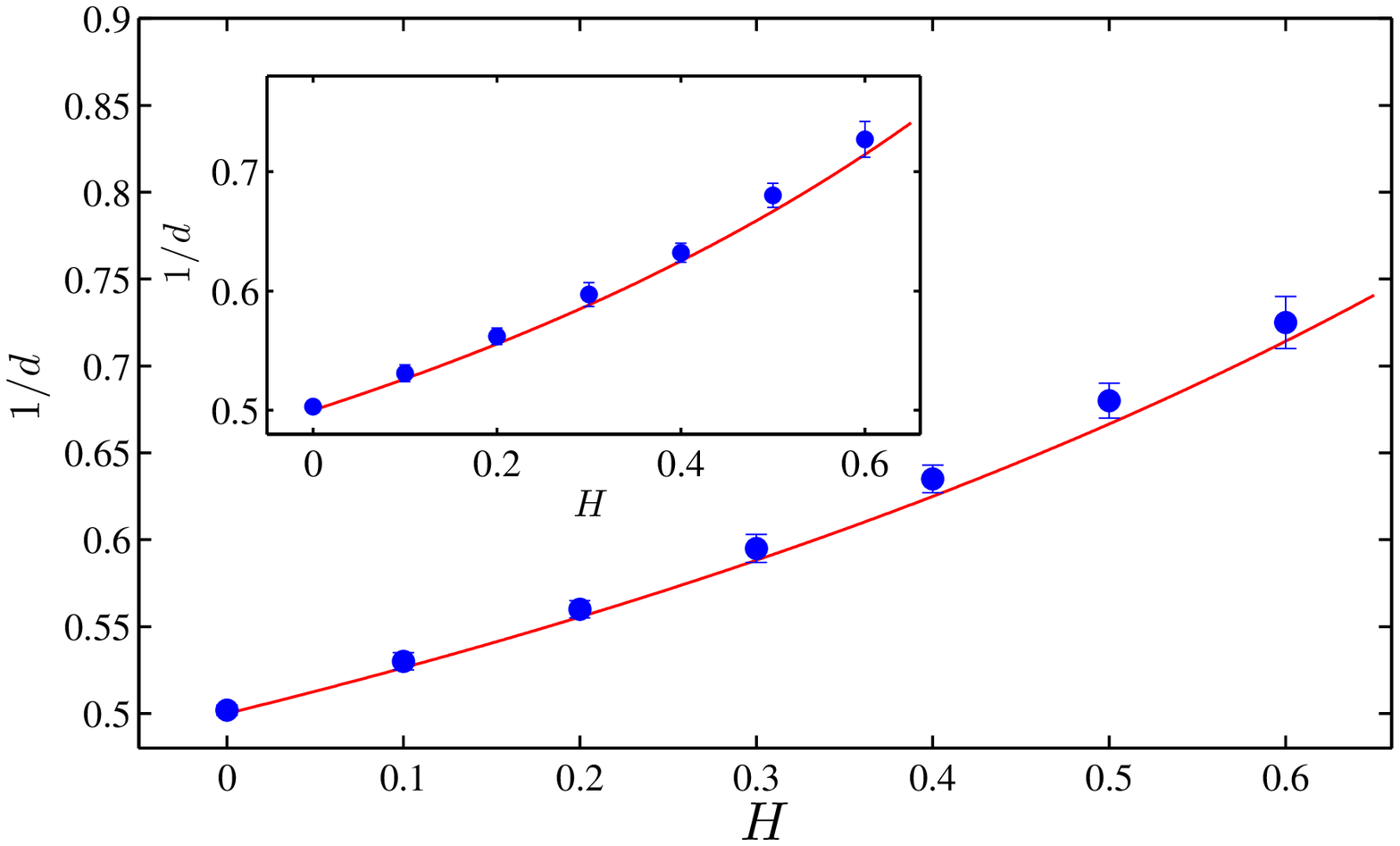}
\includegraphics[angle=0,scale=0.35]{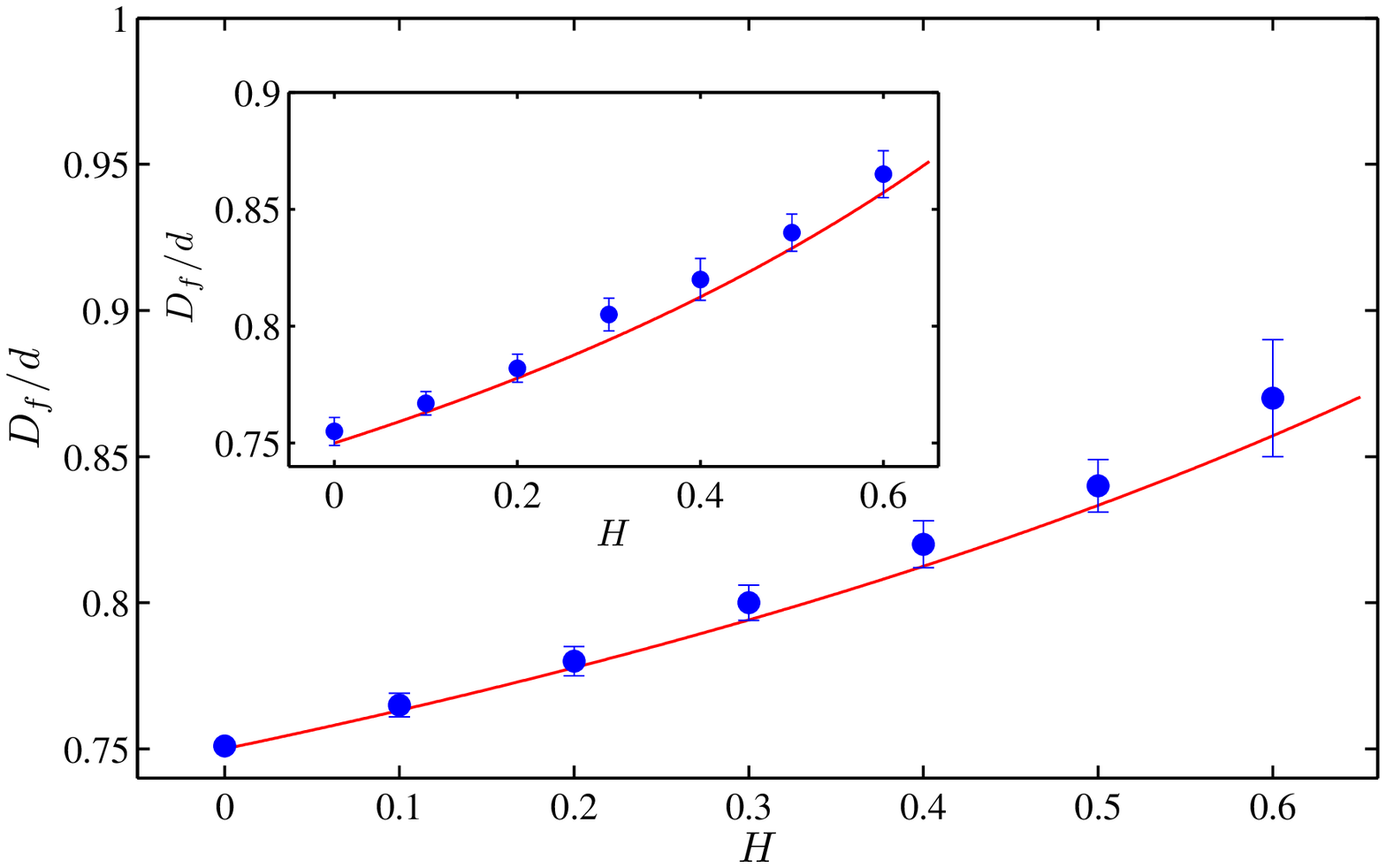}
\includegraphics[angle=0,scale=0.35]{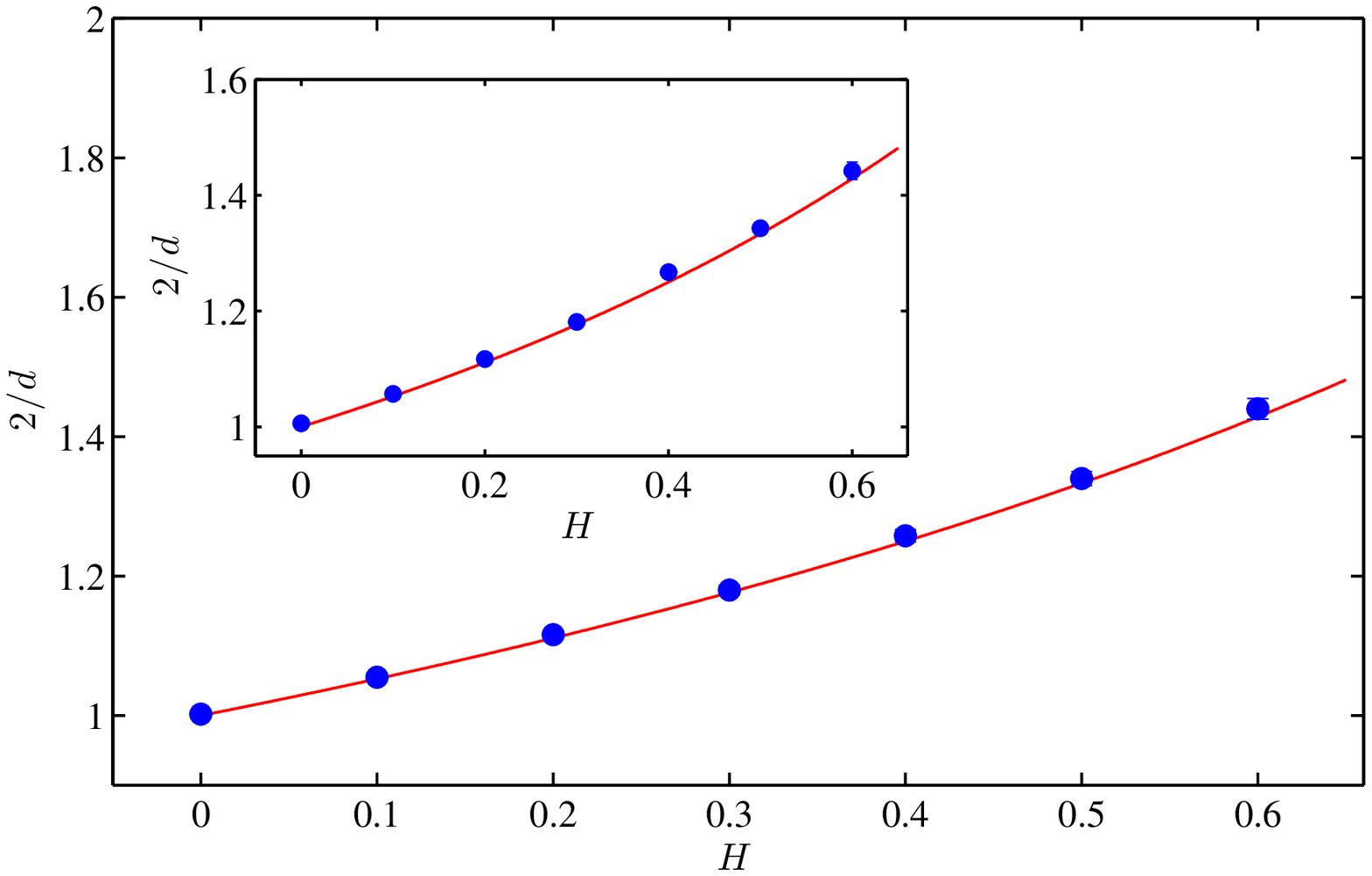}
\caption{(Color online) The numerical estimation for the scaling exponents: top $1/d$, middle $D_f/d$ and bottom $2/d$; for the WM rough surfaces with $G(x)=1-e^{ix}$ and the system size $L=2048$. The numerical computation of each exponents for the rough surfaces with $G(x)=\sin^{-1}(\sin(x))$ are shown in the insets. The solid lines show the theoretical relation between $1/d$, $D_f/d$ and $2/d$ with the \textit{Hurst} exponent $H$.}
\end{center}
\label{Figure:11}
\end{figure}

Zipf's law is one of the most  prominent examples of power laws in nature \cite{ RV,Mandel,Mandel2, JST, Zipf}. If we assign ranks to a measurable quantity $\mathcal M_n$ according to their size, there is an interesting power law relation in the size-rank distribution which is called by Mandelbrot the Zipf's law. In this method the greatest one measured quantity has rank 1 ($\mathcal M_1$), smaller rank 2 ($\mathcal M_2$) and so on. Zipf's law states that the ranked size $\mathcal M_n$ falls off as
\begin{eqnarray}\label{Zipf_law}
\mathcal M_n \sim n^{-\gamma},
\end{eqnarray}
where $\gamma$ is a scaling exponent. The scaling laws (\ref{Zipf_law})  for  $\mathcal M_n=\lbrace l_n, A_n, R_n \rbrace$ with $l_n$, $A_n$ and $R_n$  are in order, the ranked perimeter, ranked area and  ranked radius of gyration, are
\begin{eqnarray}\label{Zipf}
l_n \sim n^{-D_f/d},\hspace{0.5cm}
A_n \sim n^{-2/d},\hspace{0.5cm} R_n \sim n^{-1/d},
\end{eqnarray}
where $D_f$ is the fractal dimension of contours and $d$ is the fractal dimension of all loops. We calculated the exponents in Eq. (\ref{Zipf}) for 2D WM rough surfaces using numerical tests. Averaged values over different realizations for each exponents in Eq. (\ref{Zipf}) as a function of $H$ is presented in Figs.~11.

\subsection{Length distribution exponent}\
The loop exponent $\tau$ can be extracted from $\tilde P(s)$ which follows  Eq. (\ref{scaling_of_P}). 
This probability distribution function at large $s$ has the statistical noise. In order to minimize these noises, we consider the cumulative distribution of loop length $\tilde P_>(s) = \int_{s}^{\infty} \tilde{P}(s^{\prime})ds^{\prime}$ with the scaling form $\tilde P_>(s) \sim s^{-\tau+1}$. The length distribution exponent $\tau$ can  be measured numerically from the power-law scaling in loop length $(10<s<1000)$. The scaling behavior of the cumulative distribution $\tilde P_>(s)$  for  $H=0.5$ is plotted in Fig.~12 for different periodic functions $G(x)$. The deviation from scaling line at large $s$ comes from finite size effects. Numerical results for the exponent $1-\tau$ are summarized in Table \ref{tab3}.

\begin{figure}[htb]
    \centering
        \includegraphics[width=0.5\textwidth]{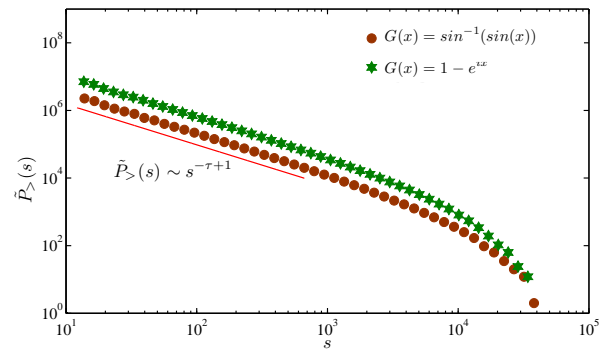}
\caption{(Color online) Cumulative loop-size distribution, for DSI rough surfaces with $H=0.5$ and system size $L=2048$. Note that this plot shows that the scaling exponent $\tau$ is independent of $G(x)$. }
\label{Figure:12}
\end{figure}

\begin{table}[htp]
\begin{center}
\begin{tabular}{|c|c|c|c|}\hline
$H$& Theory &  $G(x)=1-e^{ix}$  & $G(x)=\sin^{-1}(\sin(x))$ \\
\hline
$0.0$ & -1.33 &$-1.33 \pm {0.01}$ &$-1.34 \pm {0.01}$    \\
\hline
$0.1$ & -1.31 &$-1.31 \pm {0.01}$ & $-1.32 \pm {0.01}$
\\ \hline
$0.2$ & -1.29 &$-1.29 \pm {0.01}$ & $-1.29 \pm {0.01}$
\\ \hline
$0.3$ & -1.26 &$-1.27 \pm {0.01}$ & $-1.26 \pm {0.01}$
\\ \hline
$0.4$ & -1.23 &$-1.24 \pm {0.01}$ & $-1.24 \pm {0.01}$\\
\hline
$0.5$ & -1.20 &$-1.21 \pm {0.01}$ & $-1.21 \pm {0.01}$\\
\hline
$0.6$ & -1.17 &$-1.18 \pm {0.01}$ & $-1.18 \pm {0.01}$\\

\hline
\end{tabular}
\end{center}
\caption{\label{tab3} The best fit values of the
exponent $1-\tau$
derived from the scaling laws of cumulative distribution $\tilde P_>(s)$. }
\end{table}

\section{Conclusion}

In this paper we first introduced  a wide range of rough surfaces with the discrete scale invariance property $(\ref{Two_WM})$. The surfaces in the limit $\gamma \to 1$ converge to the Gaussian mono fractal rough surfaces. The contour lines of the introduced curves show clear DSI property. The fractal dimension of the contours follow the behavior of the fixed point, i.e. Brownian sheet. The numerical calculation shows that the relation $x_{l}=\frac{1}{2}$ is superuniversal which means that it is independent of $G(x)$, $H$ and $\gamma$. Insteed the exponent $n$ changes linearly with respect to the roughness exponent $H$. We also checked numerically the consistency scaling exponent relations, i.e. (\ref{hyperscaling})  and $(\ref{sum_rule})$. There are different methods to define WM rough surfaces in two dimensions, for a different example see \cite{Falconer}, we strongly believe that the same scaling relations also hold for these surfaces. Although the relations that we introduced in section ~3 are valid for almost all the mono fractal surfaces the careful studies of the multi fractal surfaces shows huge discrepancies \cite{HVMR}.

Our calculation shows that it is possible to study the DSI version of the contour lines of the scale invariant rough surfaces. 
The DSI property does not change the fractal dimension of the curves. It seems that this is a very general property \cite{GR} and possibly can be seen in a wide range of statistical models. In other words it should be possible to change the usual statistical models in  a particular way to have DSI property with  the similar fractal properties as the ordinary statistical mechanics models. 

Up to know the only difference of our models with the scale invariant rough surfaces is the DSI property. It is quite natural to try to find some other quantities that can distinguish the  continuous scale invariant rough surfaces from the discrete scale invariant rough surfaces. Since most of the well-known statistical models at the critical point show universality and in some cases conformal invariance it seems that we need a better understanding of discrete scale invariant field theories.

\begin{center}
{\bf ACKNOWLEDGMENTS}

M. G. Nezhadhaghighi would like to thank S. Hosseinabadi for many helpful
discussions, and also  H. Hadipour and R. Mozaffari for computer supports.

\end{center}

\end{document}